\documentclass[graybox]{svmult}

\usepackage{type1cm}        
\usepackage{braket}                            
\usepackage{makeidx}         
\usepackage{graphicx}        
\usepackage{multicol}        
\usepackage[bottom]{footmisc}

\usepackage{newtxtext}       %
\usepackage{newtxmath}       
\usepackage{natbib}
\usepackage{url}

\makeindex             

\begin{document}

\title*{That Damned Equation.\\ Rigour, Credit Attribution, and the Wheeler-DeWitt Equation 1962-1967}
\titlerunning{That Damned Equation} 

\author{Alexander S. Blum, Dean Rickles, and Karim Th\'{e}bault}
\institute{Alexander S. Blum \at Munich Center for Mathematical Philosophy, \email{alexander.blum@lmu.de}
\and Dean Rickles \at University of Sydney, \email{dean.rickles@sydney.edu.au}
\and Karim Th\'{e}bault \at  University of Bristol , \email{karim.thebault@bristol.ac.uk}}

\maketitle

\abstract{The notion of rigour relevant to the practice of physics is an endogenous one. Theoretical physics has its own internal norms about mathematical practice and notions of legitimate derivations or formal objects. These norms are often implicit, local, and change in substantial ways over time. Moreover, norms of rigour in theoretical physics, at least in the mid-twentieth century, are primarily focused on the goal of removing barriers for concrete calculation and clear conceptualisation. Rather than pursuit of rigour for its own sake, or to achieve some `higher' standard of truth, theoretical physicists seek to `rigorise' initial, semi-formal constructions in order to render formally well-defined models with which they can make concrete contact with the world, through calculations of quantities of interest. In what follows we will support this thesis based upon a detailed historical case study of the development of and attribution of credit for the Wheeler-DeWitt equation in the period 1962-67. Drawing upon archival material and the published record we develop and defend the explanatory hypothesis that it was the rigorisation of the equation via the expression for the inner product that was the crucial step in the work of Wheeler and DeWitt rather than the statement of the equation itself. }

\section{Rigour and Reality}
\label{sec:1}

\subsection{Between Mathematics and Experiment}
\noindent Jean Gaston Darboux's opinion on examining Poincar\'{e}'s  1878 doctoral thesis provides us with a useful way into the issues of this chapter:

\begin{quote}
At first glance, it seemed clear to me that it [Poincar\'{e}'s thesis] was out of the ordinary and fully deserved to be accepted. Certainly it contained enough results to furnish material for several good  theses. But it must be said without hesitation if an accurate idea is to be of the way in which, Poincar\'{e} worked, many points required corrections and explanations. Poincar\'{e} was an intuitive mathematician. Once at the summit, he never retraced his steps. He was content to have broken through the difficulties, and left it for others to trouble themselves with mapping out the royal roads that would lead more easily to the goal. He willingly did the corrections and tidying which seemed necessary to me. But when asked him to do so, he explained to me that he had many other ideas in his head; he was already occupied with some of the great problems whose solution he was to give us.\footnote{As cited in \citep{goroff}.}
\end{quote}

\noindent This description of Poincar\'{e}'s style could  be applied just as well to the style of many physicists' work, where the intuitive results achieved have much in common with what has been labelled ``theoretical mathematics'' or ``speculative mathematics'' \citep{jaffe}. But the process is always two-step in mathematics, according to Jaffe and Quinn: one can make flights of the imagination from pure intuition, but the  proof-oriented procedures of ``rigorous mathematics'' must come in to evaluate the harvest.\footnote{This is by no means universal in mathematics. In his later years at least, Alexander Grothendieck expressed the idea that it was the intuitive leap that really mattered and, as Ian Hacking expresses it, ``all the long, boring and non-memorable proofs are just so many crutches for the halt and lame'' \cite[p. 30]{hack}.} Indeed, Jaffe and Quinn model this view after theoretical physics, which also explores structures in a free manner, but must be backed up by experimental/empirical confirmation: Proof is to ``theoretical mathematics'' as experiment is to theoretical physics. But these two strands can become curiously entangled, especially in the context of quantum gravity research. In such cases, making the vision more rigorous might threaten its physical applicability, yet there is not enough physical evidence to support the vision either, so such work falls in a strange limbo state between abstract and concrete. 

In a now famous study of the birth of the modern scientific method, \emph{Leviathan and the Air-Pump: Hobbes, Boyle, and the Experimental Life} \citep{shapshaf}, Steven Shapin and Simon Schaffer uncovered a seemingly simple disagreement between Robert Boyle and Thomas Hobbes over the proper road to worldly truths: experiment (Boyle's best bet) versus mathematical principles (Hobbes' best bet). For Hobbes, there were simply too many ways in which experimental errors could creep in, so as to confound any result; unlike mathematical reasoning, as best exemplified in geometry. Of course, the subsequent founding of the Royal Society, which positioned experiment  as the essential core of science, would win the day---for socio-political reasons, as much as `Truth' with a capital T, for Shapin and Schaffer.\footnote{There was something \emph{ungentlemanly} about restricting knowledge to  conclave of experts claiming to know its invisible, arcane rules: knowledge should be a club, open to all and reproducible to and by anyone with a decent set of sensory faculties.} 

Isaac Newton's \emph{Principia} established standards for physics that were previously allotted only to mathematics and logic, while  staying on the right side of the empirical code of conduct of The Royal Society. He was seemingly able to present empirical science as an exact science, and thus reinstate the value of mathematical reasoning in natural philosophy. This embrace between Hobbes' and Boyle's bets can best be seen in his Optical Papers of 1672---see \cite{shap}:

\begin{quote}
    I therefore urge geometers to investigate nature more rigorously, and those devoted to natural science to learn geometry first. Hence the former shall not entirely spend their time in speculations of no value to human life, nor shall the latter, while working assiduously with an absurd method, perpetually fail to reach their goal. But truly with the help of philosophical geometers and geometrical philosophers, instead of the conjectures and probabilities that are blazoned about everywhere, we shall finally achieve a science of nature supported by the highest evidence.
\end{quote}

\noindent Over the course of the 19th and early 20th century, modern theoretical physics emerged as a compromise and a mediation between the extremes of mathematics and experimental (``Baconian'') physics \citep{kuhn, warwick, jungnickel}. But even today, this ideal harmonization -- where mathematical rigour is seamlessly connected to the experimental life -- is not always achieved. In the search for a quantum theory of gravity, in particular, \emph{both} mathematical rigour and experimental confirmation remain elusive. Physical concepts and frameworks are often just \emph{intuited}, because there is not yet a rigorous mathematical description available (we turn to this problem in the subsequent sections). Conversely, experimental confirmation of quantum gravity theories is difficult to achieve for reasons of inaccessibility. Quantum gravity is new, and scientific methodology is old. So we might have to reassess, or at least examine more closely, what are taken to be \emph{givens} in scientific research, and in the different areas of research rather than presupposing a one-size-fits-all approach. Here we do that with respect to the precise role (or roles) of rigour in a specific, difficult example from quantum gravity research---the Wheeler-DeWitt equation---using the more famous example of the Dirac delta function as a comparison case. 

The reader may well be most familiar with these issues from the debates on string theory. Here, the decoupling between experiment and mathematics is manifest, with the mathematical side of things facing exactly Newton's charge against the pure geometers, that  its claims constitute ``speculations of no value to human life'' (that is,  no empirical substance). But this disconnect from the world of experiment applies to quantum gravity more generally. 

In the case of string theory it has been argued that the lack of novel experimental predictions is not in fact a problem for determining the veracity of the theory, as a theory of the world. The idea is that the mathematical structure is so tightly constrained (including by `old evidence' from known particle physics and gravitational physics) that there is simply no other theory that is possible, and so it must be the actual physical theory of our world---\emph{cf}. \cite{dawid:2013} for a philosopher's defence of this view. In this case, rigour becomes a pivotal criterion for success, and string theory thus strongly aligned itself with pure mathematics, even leading to entirely new mathematical discoveries \citep{katz:2006}. But this seems to be moving away from the harmonious integration of mathematics and experiment, and a lot of the criticism of string theory can certainly be understood in this sense.  

Now one might expect that loss of contact with experiment would inevitably lead to an increased reliance on rigour and a reversion to a Hobbesian view of fundamental physics. We might thus expect to find developments similar to those in string theory throughout the history of quantum gravity. Rigorous results have indeed played an important role in shaping this research field more widely, such as the non-renormalizability proofs of the 1970s and 80s, which formulated a challenge that any theory of quantum gravity will have to answer. But we also have the strange case of the Wheeler-DeWitt equation. It became the defining equation of the canonical approach to quantum gravity, which was initially based on the traditional quantization methods of early quantum mechanics, rather than on the covariant methods of later particle physics. Yet, the Wheeler-DeWitt equation was not mathematically well-defined as originally stated. And this is still the case, more than half a century after its proposal in the 1960s: it has been described as ``ill-defined'' and giving ``meaningless results'' \cite{Rovelli:2015} or as ``a highly singular functional differential equation, which most likely cannot be made mathematically well defined in this form'' \citep[p. R201]{nicolai:2005}.\footnote{The focus of our analysis is on the rigour problems for the Wheeler-DeWitt equation as they were understood in the 1960s. We will, however, draw upon insights from the modern mathematical and theoretical physics literature -- such as the papers cited above, \cite{landsman:1995} and \cite{Thiemann:2007,thiemann:2007b} -- when relevant for clarification. Another issue with the Wheeler-DeWitt equation is that it remains highly abstract, in particular as concerns the role of time. We will not focus our discussion on the  `Problem of Time' although there is a large literature on the subject. This includes various criticisms of the standard argument towards the frozen formalism which we neglect in what follows due to constraints of space (and time). The intersection  between the more formal problems of rigour and the more conceptual aspects of the problem of time would be a suitable topic for further historical and philosophical analysis.  See \cite{Kuchar:1991,isham:1993,pons:2010,anderson:2017,Gryb:2016a,gryb:2024,CASADIO2024169783}.} Moreover, the status of the equation can hardly be seen to stabilised by appeal to experiments in the regime in which it is expected to be relevant (since such experiments are plausibly beyond even distant human capacities). The leads to the natural, and yet strangely under-explored, question of what precisely we mean by `rigour' in the context of theoretical physics in general, and the Wheeler-DeWitt equation in particular. 

\subsection{Exogenous and Endogenous Rigour}

When studying the notion of rigour in the physical sciences, a first default assumption might be to simply import the concept of rigour from mathematics, where it is well articulated. This would be an approach in which the notion of rigour relevant to the practice of physics is an exogenous one. Rigour in physics just is rigour in mathematics. Rigorisation is then a process in which the intuitive or heuristic formal expressions of physicists are progressively converted into objects that are well-defined by the standards of mathematics. We will analyse this exogenous notion of rigour in more detail in Sections \ref{Sec:2} and \ref{Sec:3}, where we conclude that it suffers from notable explanatory defects, both in general and in the particular case of the Wheeler-DeWitt equation. In particular, it does not provide us with basis for understanding when and why later rigorisation is possible. We will discuss several canonical examples of such mathematical rigourisation (all due to Paul Dirac) and conclude two things: (a) If we take the exogenous notion of rigour as our standard, the success of Dirac's heuristic methods remains entirely unexplained, appearing as ``rigourous magic.'' (b) Since in the case of the Wheeler-DeWitt equation it has not been possible so far (and it seems unlikely that this will happen any time soon) to convert it into an expression that is anywhere close to being well-defined by the standards of mathematics, the exogenous notion of rigour suggests that the Wheeler-DeWitt equation is either a lost cause or should simply not be judged by standards of rigour at all.

The central thesis of this chapter is that the notion of rigour relevant to the practice of physics is actually an endogenous one. Theoretical physics has its own internal norms about mathematical practice and notions of legitimate derivations or formal objects. These norms are often implicit, local, and change in substantial ways over time. Moreover, norms of rigour in theoretical physics, at least in the mid-twentieth century, are primarily focused on the goal of removing barriers for concrete calculation and clear conceptualisation. Rather than pursuit of rigour for its own sake, or to achieve some `higher' standard of truth, theoretical physicists seek to `rigourise' initial, semi-formal constructions in order to render formally well-defined models with which they can make concrete contact with the world, through calculations of quantities of interest. When we base our analysis on this endogenous notion of rigour, the difficulties associated with rigorous magic disappear: (a) There is no more mystery as to why Dirac's heuristic methods worked, as they enabled calculation and conceptualization and were thus up to the standards of endogenous physicists' rigour -- it is the later mathmematical rigourisation that now becomes tangential. (b) We can now apply standards of rigour also to more problematic cases, and we find, again using the endogenous notion of rigour, that the Wheeler-DeWitt equation actually represents a rigourisation of earlier, more formal-heuristic attempts at setting up a Schr\"{o}dinger equation for general relativity. This will also resolve the historiographical problem of credit attribution: while others had (or could easily have) written down equations that looked very similar (in particular in hindsight), it was only through the work of Wheeler and DeWitt that this equation at least started to meet standards of endogenous rigour.

We have mentioned two elements of endogenous rigour: allowing for conceptualization and enabling calculations. Interestingly, these two aspects map nicely onto the equation's two namesakes. For John Wheeler, the equation was supposed to embody his intuitive, physical (ontological) vision  of what quantum gravity should be \emph{about}, namely a space of 3-geometries functioning as a possibility space for topological fluctuations of spacetime. For Bryce DeWitt it was a more formal exercise, implementing specific, and yet strictly still mathematically ambiguous, prescriptions for the inner product and operator ordering in quantum gravity, which would allow for some first simple calculations in quantum cosmology. Note that the derivation of the equation's general functional form was \textit{not} a significant contribution by either DeWitt or Wheeler. The functional differential equation expression for the Wheeler-DeWitt equation first published in DeWitt's 1967 paper and presented at various talks by DeWitt and Wheeler in the years previous was \textit{no great technical innovation}. It is essentially already contained in the Hamilton-Jacobi equation for general relativity  written by Peres in 1962. 

Wheeler and DeWitt should not have been and were not at the time given much credit for simply stating the equation. Rather, we propose an explanatory hypothesis of both the development of and attribution of credit for the Wheeler-DeWitt equation in the period 1962-67 that can  best be understood by appeal to the notion of theoretical physicists' rigour. It is the rigourisation of the equation according to the contemporary standards of theoretical physics, and not mathematical physics, that DeWitt and Wheeler were pursuing and were given credit for through the convention of naming the equation after them.\footnote{By way of contrast, we might  mention Richard Feynman's path-integral (functional integration) formulation of quantum mechanics and quantum field theory (worked out under John Wheeler's supervision, as a PhD thesis), which began without a firm understanding of the integration measure. Again, there was no mathematical definition of what was being used as a kind of tool for making calculations. Indeed, it is more serious than a simple lapse of rigour: the Lebesgue measure technology required to define the volume element for the kind of infinite-dimensional, non-locally compact, function space forming the domain of integration for the path integrals  breaks down. But here Feynman's tool was \emph{already} capable of making concrete predictions (e.g. Feynman's own relativistic computation of the Lamb shift) and so satisfied the theoretical physicists' standard of rigour without further rigorisation required \emph{in this particular context}. The problem of rendering the framework rigorous (in the sense of being shown to correspond to well-defined mathematical structures) was initially tackled by C\'{e}cile DeWitt-Morette \citep{cecile}, and then in far more detail over several decades, culminating in \citep{ceccar}---there were also other routes to mathematical taming, such as lattice approaches that discretize spacetime to achieve rigour, but these distort the original approach. But the crucial point here is that any subsequent rigorisation was not part of the theoretical physicists' world, but was a case of exogenous rigorisation for and by the mathematical physics community.}  This rigorisation was achieved via Wheeler's ontological interpretation combined with DeWitt's prescriptions for the inner product and operator ordering. Our alternative history of the Wheeler-DeWitt equation contrasts with standard historical accounts of the development of the Wheeler-DeWitt equation, including those given by DeWitt and Wheeler themselves in later recollections. We will show, however, that it is supported not only by its greater explanatory virtue but by contemporary documentary evidence from the unpublished record.  

Before we come to this, let us step back and consider in more detail the idea of rigour in the theoretical physics context. We begin by re-examining the exogenous notion of rigour and find, in Section \ref{Sec:2}, that it can sensibly applied only to instances of `rigourous magic'. We discuss several such instances in Section \ref{Sec:3}, but then transition to a case where this category cannot be applied, the Wheeler-DeWitt equation, which will then be our focus for the main part of the paper.

\section{Rigorous Magic}
\label{Sec:2}

What do we mean by ``rigour'' in the context of theoretical physics? The most simple and direct approach is to assume that the notion of rigour in theoretical physics is the same as the notion of rigour in mathematics. This, in turn, is standardly defined via the notion of formalisation in the context of proof. For example, what might be taken to be the \textit{Standard View of Rigour} runs something like the following (cf. \cite{Weatherall:2026} :
\begin{quote} 
 Mathematical proof is the primary form of justification of mathematical knowledge. But in order to count as a proper mathematical proof, and thereby to function properly as a justification for a piece
of mathematical knowledge, a mathematical proof must be rigorous [..] a mathematical
proof P is rigorous if and only if P can be routinely translated into a formal proof' \citep[pp. 209-10]{hamami:2022} 
\end{quote}
In this context, the `rigorisation' problem for physicists and mathematicians can be understood to consist in the same problem of `repairing' or `cleaning-up' mathematically under par derivations such that they become genuine mathematical proofs in the sense of the Standard View. Clearly, such problems are very common in physics, with a particularly famous case being the outstanding problem of `rigorisation' of quantum field theory \citep{blum:2025}. Still, according to the standard view, such rigorisation activity is  continuous with similar work found in the context of the history of mathematics proper. 

However, what guarantees that the work of ``intuitive mathematicians'' is in fact amenable to rigorisation? It seems that we have an explanatory problem with regard to how physicists' persistent non-rigorous use of mathematics is successful to an almost magical degree. To echo Eugene Wigner, we have a case of \emph{really} unreasonable effectiveness of mathematics here, which now includes unrigorous (relative to standards external to theoretical physics) mathematics. If we are to believe that theoretical physics should indeed be judged by the exogenous standard of rigour supplied by mathematics, we are forced to view much of the work done in theoretical physics as \textit{rigorous magic}, i.e., as purely formal expressions that are not mathematically well-defined, and that we can fully understand only in hindsight, once they are translated into unambiguous expressions in terms of \textit{well-defined} mathematical objects (by the exogenous standards of rigour of mathematics).

In order to make this discussion less abstract, we will in the following section consider three examples where this later mathematical rigourisation indeed happened. All three examples are taken from the work of Paul Dirac, who has, on the basis of these and other examples, indeed garnered somewhat of a reputation as a physico-mathematical wizard. We will be labelling these episodes as rigorous magic, but this is a purely historiographical category: a heuristic calculation by physicists that is later \emph{formally precisified} by dedicated mathematical work. Indeed, we do not think that it can serve as an explanatory category. One might, in principle, consider two hypotheses as to how rigorous magic works. First is the idea, already implicit above, of \textit{mathematical intuition}. The idea is to suggest that the great physicist-magicians (Dirac or Feynman) are ``intuitive mathematicians'' who provide derivations and introduce new formal objects in just such a way that `rigorisation' in terms of translation into rigorous proof and well-defined objects is always possible. Second, and perhaps more plausibly, we might suggest that, in fact, it is not always the case that there is a mathematically rigorous version of everything physicists do, it is just that we ignore the historical cases where this doesn't work out or isn't even tried. Thus the appearance of `rigorous magic' is a product of confirmation bias.  

 The problem with either explanation is that they make interpretation of the actual mathematical work of theoretical physicists heavily dependent on whether mathematicians (or mathematical physicists) revisit the same problem later and are successful in their rigorisation work. However, \textit{physicists typically do not actually care about whether this happens later} and when they do criticise derivations or formal objects in theoretical physics it is not because there is a lack of rigour in the mathematical sense or a promissory note that this is achievable. Rather, the problem the physicists typically have with lack of rigour is that it places obstacles to clear physical conceptualisation and calculation within concrete models. Correspondingly, the goal in rigorisation for physicists is formal precisification such that these obstacles are removed, not such that unambiguous expression by the standards of mathematics is obtained. From the perspective of methodology of science, the problem of explaining rigorous magic is simply the wrong one  precisely because it focuses on the mathematicians' notion of rigour rather than that relevant to physicists. Moreover, since theoretical physics' norms of rigour  are historically situated and pliable, their achievability from a semi-formal heuristic starting point is hardly miraculous. In what follows, we will consider three examples of rigourous magic in development of our argumentative foil, before we turn to the exogenous notion of rigour and its fruitful application to the case of the Wheeler-DeWitt equation.

\section{Dirac's Magic}
\label{Sec:3}

Paul Adrian Maurice Dirac was a pioneer of startlingly insightful and yet not entirely mathematically rigorous physics. The delta function that bears his name is the most notable singular example and his famous bra-ket notation for expressing the quantum formalism is perhaps the most familiar. In each of these two cases we have an example of \textit{Rigorous Magic}. Each are heuristic semi-formal expressions for objects that are not mathematically well-defined as presented but were later \textit{formally precisified}: i.e. translated into an unambiguous expression in terms of \textit{well-defined} mathematical objects (by the standards of mathematics), before turning to our third example, which is is ultimately most relevant to our discussion: canonical constraint quantization or Dirac quantization. While constraint quantization is arguably also an example of rigorous magic, it is also the context in which the Wheeler-DeWitt equation is developed, which decidedly does not fit into this historiographical category.

As noted by \cite{sep-qt-nvd} the delta function was used by various physicists and mathematicians long before Dirac. For example, its close relative the Heaviside step function was introduced in the late nineteenth century. However, the widespread use and naming of the function within physics can be traced to the effective use put to it for quantum theory within Dirac's famous textbook \citep{Dirac:1930}. The formally problematic aspect of Dirac's treatment is specifically criticised in otherwise laudatory remarks in von Neumann's own textbook on the topic in which he notes that the delta function is an ``improper function with self-contradictory properties'' and might be understood as a ``mathematical fiction''. However, although the Dirac delta function was ill-defined in Dirac's usage, it can be given a formal precisification via Schwartz’s  theory of distributions \citep{sep-qt-nvd}. This fits with the standard account of mathematical rigour in physics where we find a trade-off between rigour and pragmatics under which the latter may trump the former so long as  one's use of ill-defined objects involves no contradiction and these objects can later be given a fully rigours expression. Thus, \cite{sep-qt-nvd} exculpate Dirac on the grounds that he is  ``of course fully aware that the [delta] function is not a well-defined expression' but he should not be in trouble since `as long as one follows the rules governing the [delta] function...then no inconsistencies will arise [and since] the [delta] function can be eliminated... it can be replaced with a well-defined mathematical expression.''
 
 The story of the bra-ket notation is closely related. In this approach the basic objects of quantum theory, quantum states, are compactly written as `kets' and their adjoints as `bras'. So for example, a position eigenstate can be written $\ket{x}$, its adjoint $\bra{x}$, and the inner-product between two eigenstates  is simply $\braket{x|y}$. This notation was first introduced by Dirac in a paper in 1939, was adopted in the 1947 third edition of his textbook and gained widespread usage thereafter. Like the delta function, Dirac's bras and kets for position and momentum eigenstates were not well formed mathematical objects even by the standards on the mathematics of the day. However, also like the delta function, they were given respectable formal precisification in later work via the theory rigged Hilbert spaces and the Gelfand triple.\footnote{Following \cite{sep-qt-nvd}: Various mathematicians contributed to the rigorisation of Dirac’s  framework. One key element was Schwartz’s theory of distributions \citep{Schwartz1945,Schwartz1951}. A second key element is the notion of a nuclear space \citep{Grothendieck1955} which made possible the generalized-eigenvector decomposition theorem for self-adjoint operators in rigged Hilbert space — for the theorem see \cite[119--127]{GelfandVilenkin1964}, for a historical account see \cite[756--760]{Berezanskii1968}.} Indeed one can easily find quotes such as: ``Rigged Hilbert spaces are well known as the structure which provides a \textit{proper mathematical meaning} to the Dirac formulation of quantum mechanics."  (italics added). Note here, that once more the issue of mathematical rigour is not with regard to standards of proof but rather the issue of a set of formal objects and expressions within a formulation of a physical theory not being unambiguously expressed according to the standards of mathematics. The have `proper mathematical meaning' is to be unambiguously statable in terms of well-defined formal objects. And, once more, we have the idea of such initial heuristic expressions later proving formally precisifiable.

Of all the many important problems that Dirac contributed to, the story of mathematical rigour is perhaps most complex, and certainly the most interesting, in the context of quantum field theory (QFT), which he pioneered in the late 1920s. The problem of rigour in QFT is a vast subject \citep{sep-qt-nvd}. Here we focus on the specific case of field theories symmetric under infinite-dimensional invariance groups, colloquially known as gauge theories. We further focus on a particular approach to the quantization of such field theories, which Dirac pioneered in the late 1940s, many years after his early work on QFT.

There are several ways to transition from a classical theory to the corresponding quantum one. What might justly be called the \textit{Royal Road of Quantization} is the canonical approach pioneered by Heisenberg, Born, Jordan, Dirac, and von Neumann, and others in the mid- to late 1920s. This road runs from promotion of Poisson bracket structure of a Hamiltonian theory to commutator structure of a quantum theory. More abstractly, this relies upon the \textit{symplectic structure} of a Hamiltonian theory with a \textit{regular} as opposed to \textit{irregular} Lagrangian -- where there are no `gauge degrees' of freedom indicated by the existence of arbitrary functions in the equations of motion. The problem of making this version of quantization fully rigorous proved a fearsome one, but has largely been achieved within the modern framework of deformation quantization in which the quantization map can be constructed as a functor \citep{rieffel:1993,landsman:1998,feintzeig:2024}. For our purposes the most important aspect of the quantization of a theory with a regular Lagrangian structure is that it comes with a phase space which is unconstrained. This means that every point  in the space corresponds to a kinematically possible instantaneous state and every sequence of such states lying along an integral curve of some Hamiltonian function corresponds to a dynamically possible model. Canonical quantization can then proceed directly.

Dirac and others had recognised early on that the existing canonical quantization procedure was ill-defined for theories with infinite-dimensional invariance groups.\footnote{Many of the details were discovered by Dirac's student, Paul Weiss.} This was first observed for the gauge group of electrodynamics, but it was soon found that the problem was also present -- and even more complex -- for the case of gravity. In the 1950s Dirac began to focus on this problem on earnest and this work culminated in a the constrained quantization algorithm that in heuristic form is known as `Dirac quantization'. The principal steps in the procedure are as follows. We first consider a gauge theory in which the Lagrangian is irregular. For such theories the Legendre map will not be onto (and hence not invertible). Consequently, the Hamiltonian formulation will have `primary' constraints $C_i(q,p)=0$ that restrict phase space to those points that lie in the image of the Legendre map.\footnote{Consistent propagation of these constrains may require us to further restrict to a sub-manifold of `secondary' (and possibly tertiary) constraints. One also distinguishes between `first-class constraints' whose Poisson brackets with all of the constraints vanish and `second-class' constraints for which that is not the case. See \cite{Dirac:1964,Henneaux:1992a,gryb:2024}.}
 
These sub-manifolds will not have symplectic structure -- rather only the weaker pre-symplectic structure where the relevant two-form has non-trivial null directions. Dirac proposed a method to build the quantum version of such theories by first quantizing the unconstrained phase space, and then second applying the constraints at the quantum level. The second `quantum reduction' stage of Dirac quantization amounts to treating the (first class) constraint functions as operators restricting the physical state vectors `$\hat{C}_i\psi=0$'. The physical Hilbert space $\mathcal{H}_{phys}$ is then constructed by considering the space of physical states. The quantum observables are understood as self-adjoint operators on $\mathcal{H}_{phys}$. Dirac quantization thus offers a strategy for re-routing the royal road of canonical quantization around the obstructions caused by irregular Lagrangians. As we have just presented it, however, it falls well short of the standards of rigour of physics, let alone mathematics. 

From a modern mathematical perspective, largely following  \citep[p. 111-2]{Thiemann:2007}, there are five particular problems of rigour in Dirac’s version of constraint quantization: 

\begin{enumerate}
\item The problem of resolving factor ordering problems in definitions of the constraints as operators (ambiguity and singularity);
\item The problem of finding self-adjoint representations of the constraint operators; 
\item The problem of satisfying requirement for the spectrum of the constraint operators to contain zero;
\item  The problem of finding a dense definition of the constraint operators (regularisation and re-normalisation);
\item The problem of defining an inner product with associated measure on the Physical Hilbert Space.
\end{enumerate}

The first, fourth, and fifth problems (and their interaction) will be the most central to our story and we will say a little about each in the specific case of the Wheeler-DeWitt equation later. The severity of these difficulties depends on the theory involved and in particular the form of the constraints functions and the algebra that they close under.\footnote{This is particularly relevant to the third problem and the case of the Hamiltonian constraints. See \cite{Giulini:1999a,DittrichThie:2006,Thiemann:2007}.}
Moreover, in the case of field theories, the factor ordering ambiguity can lead to the existence of singularities that can only be removed by a regularisation of the relevant operator. Thus, the first  problem of factor ordering and the fourth problem of regularisation and re-normalisation become entwined. Indeed, there is an important sense in which the various problems of rigour in Dirac quantization are tokens of more general forms of rigour problems that can be identified in quantum field theory in general. We will return to this connection later. For the time being, it is worth noting that at least for some theories under some conditions, the precisification of Dirac quantization has proved the basis for a rigorous quantization, see e.g.  \cite{Giulini:1999a}. We thus have (yet another!) example of Dirac performing rigorous magic in the sense discussed. However, this is no longer true for the next step in the application of Dirac's constraint quantization to general relativity, the Wheeler-DeWitt equation.

\section{A Schr\"{o}dinger Equation for General Relativity}
\label{Sec:4}

During the 1950s there were three interconnected groups that aimed to quantize general relativity by using constrained Hamiltonian dynamics and the Dirac quantization algorithm. The interactions between them is of relevance to the later dispute regarding the derivation of the Wheeler DeWitt equation.\footnote{The early history of the canonical or Hamiltonian theory of general relativity starts with Rosenfeld's pioneering work on what is now called the constrained Hamiltonian approach in the context of an attempt to quantize linearised gravity  \citep{rosenfeld:1930}. The significance of Rosenfeld's work has only been relatively recently rediscovered and has received an excellent account in Donald Salisbury's works: \citep{salisbury:2009,salisbury:2010,salisbury:2017}.} First, there is, of course, Dirac himself. Dirac's interest in constraint quantization was originally geared towards reconciling quantum theory and \emph{special} relativity, a problem that had occupied him since the 1920s \citep{blumsalisbury:2018}. His methods were then adapted to the quantization of \emph{general} relativity by Alfred Schild and Felix Pirani, who published a couple of papers on the topic in the early 1950s. Dirac himself first tackled this problem in an important cluster of papers published between 1958 and 1959. The 1958 papers \textit{Generalized Hamiltonian Dynamics} \citep{Dirac:1958a} and \textit{The Theory of Gravitation in Hamiltonian Form} \citep{Dirac:1958b} respectively present the classical constraint algorithm and the canonical theory of gravity. In \textit{Fixation of Coordinates in the Hamiltonian Theory of Gravitation} \citep{Dirac:1959} he then considers a version of the theory in a particularly important gauge, cf. \cite{York:1972,Gomes:2011}, and provides probably the first, heuristic $\hat{H} \Psi =0$ expression for quantum gravity. Dirac's focus on quantum gravity proved short-lived. He had long been sceptical of the renormalization program in quantum field theory; and when he realized that similar methods (such as regularizations of the products of  singular distributions, which he considered ad hoc and prone to inconsistency) would have to be used in quantum gravity, he largely abandoned the project \citep[p. 543]{dirac_1968_the-quantization}. His  Yeshiva lectures \citep{Dirac:1964}, which were for many years the standard textbook treatment of constraint  quantization, do not include discussion of quantization in the gravitational case.

The second grouping is somewhat broad and consists of a Harvard-Princeton network centred on John Wheeler and including Bryce DeWitt and the famous ADM collaboration of Richard Arnowitt, Stanley Deser, and Charles Misner.\footnote{Deser and Arnowitt (like Bryce DeWitt) studied with Schwinger at Harvard, while Misner studied with Wheeler at Princeton.} Wheeler and DeWitt we will consider in more detail in Sections \ref{Sec:5} and \ref{Sec:6}. The work of ADM is densely clustered about the years 1959 to 1962 and includes  what was to become the standard formulation of canonical gravity \citep{ADMII}\footnote{Defined via an embedding of a three-dimensional hypersurface into a four-dimensional spacetime, with the canonical data defined on the 3-surfaces and curves on the constraint surface within phase space corresponding to spacetimes as represented in a 3+1 formalism.}; the first definition of energy within the asymptomatically flat case \citep{ADMII}, and the now standard reference for the topic \citep{ADMReview}. Their attitude towards the quantum theory was primarily drawn from the analogy to quantum electrodynamics (QED) and the influence of Schwinger. At the time Schwinger was promoting the use of the radiation (or Coulomb) gauge in QED, where the unphysical degrees of freedom (photons with longitudinal or time-like polarization) were eliminated from the Hamiltonian through a specific choice of gauge \citep{blumforthcoming}. In QED, this elimination could be performed before or after quantization; ADM wanted to replicate this method in quantum gravity. This worked fine in a linearized theory of gravity, which is very similar to QED \citep{ADI}; but they ultimately found themselves unable to implement their gravitational analogue of the radiation gauge in the full non-linear quantum theory.\footnote{See \citep[pp. 1600-1601]{ADMIII}. ADM gave several distinct reasons for this failure: (a) ADM considered Lorentz transformations to be ``an operation that must be allowed for any sensible quantum theory.'' However, a Lorentz transformation doesn't just involve the three-metric, it also alters the defining equations for lapse and shift. So those equations, which might naively be taken to be unquantized relations fixing the classical values of lapse and shift, need to be consistently integrated into the quantum theory, and it was not clear how to do this in the full non-linear theory. (b) At the quantum level, one loses the characteristic general covariance of general relativity that allows switching back and forth between arbitrary coordinate systems, a difficulty also emphasized more recently by \citet[p. 17]{anastopoulos_2013_a-master}. It therefore makes a difference whether the gravitational radiation gauge is imposed before or after quantization, and ADM concluded that it would lead to too many ambiguities if the gauge were already fixed at the classical level. They announced that one should thus ``try to repeat [their classical elimination of the unphysical degrees of freedom] within the framework of quantum theory,'' but ultimately did not pursue this avenue. (c) They also mentioned the factor-ordering ambiguity, which we will return to later, but clearly considered this less detrimental than the other two difficulties. One of us (ASB) would like to thank Maaneli Derakhshani for a very helpful discussion of these points, many years ago.} 

The third principal group is the Syracuse school founded by Peter Bergmann. Bergmann had been a research assistant of Einstein and worked consistently with various co-authors on the problem of canonical gravity and quantization from the late 1940s to the early 1970s. Notable papers include: \cite{bergmann:1950,bergmann:1953,bergmann:1956,Komar:1958,Bergmann:1960,Bergmann:1961,Bergmann:1966}. As argued in detail by \cite{salisbury:2022}, a particular point of contention between the Syracuse school and the approach pursued by Wheeler and DeWitt relates to the the formal and conceptual status of four-dimensional diffeomorphism invariance. While the Syracuse school considered such invariance an essential element of a quantum theory of gravity, this was far less important to the other groups discussed here. It was this contrast that, plausibly, directed the Syracuse school in a different direction from the fairly straightforward path to the functional form of the Wheeler-DeWitt equation, which does not display four-dimensional diffeomorphism invariance.\footnote{See in particular the discussion of the work of Bergmann and Komar in \citep[\S 6]{salisbury:2022}.}

The final crucial figure in this story is the Israeli physicist Asher Peres. Peres was the first to extend Dirac's work from a Hamiltonian formulation of general relativity to a Hamilton-Jacobi formulation published in \citep{Peres1962}. This work built on earlier work by \cite{Higgs:1958} that was completed during a visit to Chappel Hill, where DeWitt was then based. Peres was a visitor at Princeton University in 1961 and Syracuse University in 1962\footnote{This is according to his CV that was available until at least April 2019 at a memorial website created by his daughter \citep{PeresCV}.} and Wheeler is thanked in the paper. We also know that Peres presented related work on the initial value problem in canonical general relativity at the Stevens meetings that were a conduit between the two school in late 1961 and early 1962.\footnote{The evidence for this comes from Dieter Brill's notebook description of Peres talk at Stevens Meeting on initial value problem. Dieter Brill (personal communication).} Peres does not directly address the question of quantization in his 1962 paper and his first published analysis of the canonical approach to quantization of the gravitational field is not until his article \citep{Peres:1968}.  

The simple observation that would have been obvious to all these figures from at least the late 1950s onwards, is that since in the canonical theory of general relativity the Hamiltonian itself takes the form of a constraint, if the Dirac quantization algorithm is na\"ively applied (without any reference to the rigour problems) then the immediate and problematic result is that we end up with a time-independent Schr\"odinger-type equation  `$\hat{H}\Psi =0$'. As noted above the schematic `$\hat{H}\Psi =0$' form of the equation appears already in Dirac's 1959 paper it can also be found in \citep{Bergmann:1966}. What is now known as the Wheeler-DeWitt equation first appeared in published form in a 1967 paper by Bryce DeWitt. It was strongly promoted by John Wheeler in the mid-60s by John Wheeler, and so, while it was initially sometimes referred to just as the DeWitt equation,\footnote{See, e.g., Wheeler's Relativity Notebook 13, entry dated 14 October 1964, p. 47. The notebook is available online at \url{https://as.amphilsoc.org/repositories/2/archival_objects/451910}.} it is now routinely credited to both physicists. As noted by Karel Kucha{\v{r}} in 1973, in its first years the equation was also known as the `Einstein-Schr\"odinger' equation:
\begin{quote}
The quantum version [\eqref{WdW}] of the super-Hamiltonian constraint was called the Einstein-Schr\"odinger equation by John Wheeler, and the Wheeler-DeWitt equation by others. Because I cannot recall any other physicists who would more strenuously object to the idea of quantizating gravity than Einstein and Schr\"odinger, and because I am not John Wheeler, I shall use the second name. (\cite{kuchar:1973} quoted in \cite[p.149]{Kiefer:2012}) 
\end{quote}
The attribution of credit for this equation to DeWitt and Wheeler was disputed --  by Peter Bergman most forcefully as we shall see shortly. DeWitt, at least, distanced himself from it. As early as 1965, i.e., before the equation was even published he remarked in a talk that ``Dirac has essentially written'' the equation and that he had only ``replaced $\pi$ by $\delta/\delta \gamma$;\footnote{Notes taken by Dieter Brill on 25 February 1965. These notes are in the personal possession of Dieter Brill and we would like to thank him for making them available to us.} in his later work he called it `that dammed equation' and said it `should be confined to the dustbin of history'. The important point for our purposes is that the Wheeler-DeWitt equation  is \textit{not} an example of rigorous magic -- the equation has \textit{still not} been rigorous formulated by standards of modern theoretical physics let alone mathematical physics.\footnote{From the modern perspective, plausibly this is simply not a well-posed problem see \cite{Thiemann:2007}} The equation derived by DeWitt is (plausibly) best understood to be inherently ill-defined in the sense that it is \textit{not formally precisifiable} -- so it really is a \textit{dammed equation}, condemned to remain forever in mathematical hell. 

What was so special about the 1967 form? As a functional differential equation it follows in a \textit{completely trivial manner} from the Einstein-Hamilton-Jacobi equation derived by \cite{Peres1962}. We can see this by considering the Einstein-Hamilton-Jacobi equation and Wheeler-DeWitt Equations side-by-side. Following \cite{sundermeyer:1982} we have that if $S[^{(3)}\mathcal{G}]$ is the Hamilton-Jacobi principle functional understood to be satisfy the spatial constraints and thus be a functional of three geometries \cite{Higgs:1958}, and $\Psi[^{(3)}\mathcal{G}]$ is a wavefunction of the space of three geometries, and DeWitt super-metric defined via
\begin{equation}
G_{ijkl} = \frac{1}{2}\sqrt{g} (g_{ik}g_{jl}+g_{il}g_{jk}-g_{ij}g_{kl})\,.
\end{equation}
then then Einstein-Hamilton-Jacobi derived in Peres's 1962 paper can be written as: 
\begin{equation}
G_{ijkl}\frac{\delta S [^{(3)}\mathcal{G}] }{\delta g_{kl}}\frac{\delta S [^{(3)}\mathcal{G}] }{\delta g_{ij}} -\sqrt{g}R=0
\end{equation}
and the Wheeler-DeWitt equation \cite[(5.15)]{DeWitt:1967} is:
\begin{equation}
\label{WdW}
\left(  G_{ijkl} \frac{\delta^2}{\delta g_{ij} \delta g_{kl}} + \sqrt{g} R \right) \Psi[^{(3)}\mathcal{G}] = 0
\end{equation}

Considering the form of the equation alone it is not at all clear why DeWitt (or Wheeler) should be attributed any credit at all. This point was forcefully made by Peter Bergmann in a letter to Wheeler dated 6th October 1965:

\begin{quote}
...I should like to take up with you in private the so-called ``DeWitt equation''. As far as I understand, and I have checked this with Jim Anderson, you have called the Schr\"odinger equation applied to general relativity by this name. As a matter of fact, I had proposed doing this in 1950, several years before Bryce entered the field of research in general relativity on a full-time basis. Approximately in 1962, according to my recollection, Bryce, I, and others discussed in New York, where he came to visit, the problem of determining factor sequences in all the constraints which would reproduce the commutator algebra of the non-quantum theory.

I have not followed through on this particular line of inquiry, because I considered that certain fundamentals have not been resolved.

I should be unhappy if a squabble about priorities should cause ruffled feelings. At the same time I feel that unless there are pertinent facts unknown to me, Bryce obtained the idea of a Schrödinger equation from me more than ten years ago and can hardly be credited with having originated it. I should be grateful to you if you would inform me of any facts at variance with this evaluation.
\end{quote}
Bergmann's letter then ends by pointing to the 1962 Peres paper -- in which Wheeler is indeed thanked. Wheeler's handwritten note on the letter notes his reactions as: 
\begin{quote}
Phoned Misner on 18th October. Said lots of people had to do with it, though none had DeWitt's inner product. He suggested acronym having to do with all who contributed: maybe I showed Misner $\Psi(^3\mathcal{G})$ for his thesis. I should check with Peter Higgs.
\end{quote}
Though somewhat cryptic,\footnote{$\Psi(^3\mathcal{G})$ is Wheeler's notation for the Schr\"{o}dinger wave function as a function of the 3-geometry. In his 1957 thesis, Misner was still taking the argument of the wave function to be a 4-metric. In a 1958 paper, written in response to Dirac's work on constraint quantization, Peter Higgs first wrote the wave function as a function of the three-metric.} this note clearly offers a degree of support for our explanatory hypothesis: according to Misner at least the inner product was what was distinctive in DeWitt's contribution. We will consider two other sources of contemporary evidence on this point later.  

The focus in the later recollections of Wheeler and DeWitt is upon the role of Peres. The narrative regarding credit attribution is, however, sharply contrasting. DeWitt's version is the one most often retold and can be found in written form as told by DeWitt in the \textit{Proceedings of the Eighth Marcel Grossman Conference}, held in Jerusalem in 1997 \citep{DeWitt:1997}.\footnote{Substantial sections of this paper are also re-printed in \cite{dewitt:2011a}} The passage key to our story runs as follows: 
\begin{quote}
John Wheeler, the \textit{perpetuum mobile} of physicists, called me one day in the early sixties. I was then at the University of North Carolina in Chapel Hill, and he told me that he would be at the Raleigh-Durham airport for two hours between planes. He asked if I could meet him there and spend a while talking quantum gravity. John was pestering everyone at the time with the question: What are the properties of the quantum mechanical state $\Psi$ and what is its domain? He had fixed in his mind that the domain must be the space of 3-geometries, and he was seeking a dynamical law for $\Psi$. 
I had recently read a paper by Asher Peres  which cast Einstein's theory into Hamilton-Jacobi form, the Hamilton-Jacobi function being a functional of 3-geometries. It was not difficult to follow the path already blazed by Schr\"odinger and write down a corresponding wave equation. This I showed to Wheeler, as well as an inner product based on the Wronskian for the functional differential wave operator. Wheeler got tremendously excited at this and began to lecture about it on every occasion. I wrote a paper on it in 1965, which didn't get published until 1967. My heart wasn't really in it [...] But I thought I should at least point out a number of intriguing features of the functional differential equation, to which no one had yet begun to devote much attention. 
\end{quote}
Wheeler's recollections similarly place emphasis on the form of the equation and the relation to the work of Peres. They are, however, strongly divergent on the importance of DeWitt. In particular, when discussing the origin of the equation in the tenth of his twenty-two interview sessions with Ford (conducted between 1993 and 1995) he makes the following remarks:
\begin{quote}
I had been so enthusiastic for the sum-over-histories way of describing quantum mechanics and the transition from quantum mechanics to classical mechanics that I couldn't help looking for a similar transition in the case of relativity. It was a great help to have Valentine Bargmann as a colleague, because he really knew relativity and analytical dynamics, and I can recall his discussing a paper of Asher Peres. Peres at that time, if I remember correctly, was at Syracuse University. Peres had a way to express the Hamilton-Jacobi form of general relativity, which was forerunner of what I later called superspace and the wave equation in superspace. Bargmann helped to clear up exactly what it was that Peres had done. People today talk of the Wheeler-DeWitt quantum equation for the dynamics of geometry, but, in view of Bargmann's inspiration, it would have been more appropriate to call it the Bargmann-Wheeler equation.
\end{quote}
In terms of the origin story and the role of the Peres-Hamilton-Jacobi equation, it is not entirely clear how to resolve the discrepancy between the two accounts. On the one hand, as noted above, we know that Peres did indeed have interactions with both the Syracuse school and Wheeler himself in the early 1960s. The Peres paper was published in October 1962 (received May 1962) and so it is very plausible that Bargmann and Wheeler would have been well aware of this work in the early '60s, independently of DeWitt. On the other hand, Wheeler's \textit{Les Houches Lectures} on quantum gravity, delivered in 1963 published in 1964, show that at this point he clearly was not aware of the implications of Peres' for possible constraints on the wave equation in superspace (i.e. the space of diffeomorphism invariant three geometries). It is only in his \textit{Battelle Rencontres Lectures}, delivered in 1967 published in 1968, that discussion of Peres' work appears. Thus, it would make sense for there to have been some relevant interaction -- whether with Bargmann, DeWitt  or both -- which made Wheeler fully aware of the connection between Peres' Hamilton-Jacobi formulation and the soon-to-be eponymous constraint on the wavefunction in superspace sometime in the mid-60s. However, as we have already noted, the mere form of the Wheeler-DeWitt equation is a rather trivial re-writing of Peres's Hamilton-Jacobi,  analogous  to that Schr\"odinger had originally used. Although the role of Peres is crucial in the recollected origin story, it is far from clear that this reflects what happened at the time, rather than a hindsight
reconstruction.

Moreover, the recollected origin story is thus particularly unclear regarding \textit{both} what it is that was novel to Wheeler or DeWitt in the derivation of the equation \textit{and} who influenced whom and at what time. In what follows we will draw upon contemporary documents to understand to isolate a plausible hypothesis of what was in fact going on in the `derivation' of the Wheeler-DeWitt equation in the mid-1960s. The hypothesis is as follows: it was the problem of defining a \textit{rigorous enough} inner-product and operator ordering that prevented physicists accepting a putative Schr\"{o}dinger equation for General Relativity as a well-formed equation according to the standards of theoretical physics of their time. DeWitt responded to this problem by finding a form of the inner-product and approach to operator ordering that met the norms of theoretical physics of that time for an adequate expression. The attribution of the equation to Wheeler and DeWitt is plausibly on that basis. This is notwithstanding the fact that the inner-product is ill-defined by standards of modern theoretical physics (e.g. lack of well-defined measure) and the operator ordering convention was soon recognised to not be uniquely well-justified.\footnote{By 1972 Misner had already taken a very different view and studied the family of Laplacian orderings in a systematic way arguing for a unique choice among these orderings based on conformal invariance \citep{misner:1972}.} 

The case of the Wheeler-DeWitt equation is not one of rigorous magic -- neither did DeWitt provide a \emph{mathematical} rigourisation of earlier heuristic work, nor was DeWitt's work later rigourised by mathematicians. Rather, it is taken to illustrate the \textit{explanatory irrelevance} of the pursuit of exogenous mathematical standards of rigour to the process of rigorisation pursued by the historical actors. By contrast our explanatory hypothesis for this case supports our central thesis that theoretical physics has its own endogenous norms about legitimate derivations or formal objects and that these norms are what motivates much work in theoretical physics. Clearly such norms may be implicit, local, and change in substantial ways over time but in least in the case in question we submit that they are closely related to goals of enabling conceptualisation and calculation in simple models. In what remains of the paper we will hope to show that by better understanding the reasoning that led Wheeler, DeWitt and the physics community more generally to accept the equation of this form at that time we can also better understand the conception of rigour in theoretical physics more generally. 

\section{The Story of the Wheeler-DeWitt Equation}

\subsection{DeWitt's Story}
\label{Sec:5}

It is somewhat ironic that Bryce DeWitt's name is now forever linked with the central equation of canonical quantum gravity. His remark, made in 1997, that the Wheeler-DeWitt equation should be ``confined to the dustbin of history'' is frequently quoted \citep[p. 58]{dewitt:2011a}. But this was hardly a late-in-life disavowal of the ambitions of his youth -- as he remarked later in the Jerusalem talk that kicks off with the dustbin quote, even at the time his ``heart wasn't really in it.'' He had pursued the canonical approach to quantum gravity in the 1950s, when it appeared to be the royal road to non-perturbative quantum gravity \citep[pp. 136--143]{blum_2017_the-1957}. But in the 1960s, he instead took up the non-perturbative covariant (functional) methods that his PhD advisor Julian Schwinger had developed over the course of the 1950s and was feverishly working on constructing a theory of quantum gravity on this basis -- as witnessed by his monumental 1963 Les Houches lectures on a ``Dynamical Theory of Groups and Fields.'' When these lecture notes were published as a book in 1965, a reviewer remarked that DeWitt's quantization was carried out ``in a manifestly covariant fashion (i.e., an overall space-time view is adopted and canonical formulations are ignored)'' \citep{mullin_1966_review}. It is thus not that DeWitt considered the Wheeler-DeWitt equation to be wrong \emph{per se}, he simply considered it to be inferior to the covariant approach to quantum gravity.\footnote{This reading was confirmed to us by Claus Kiefer, who reported on a conversation he had with DeWitt on the topic in Santa Fe in 1990 (Email from Claus Kiefer to the authors of 24 April 2026).}

The details of DeWitt's covariant program need not concern us here. But one thing is important to highlight: DeWitt showed a keen awareness of the pitfalls involved in attempting to give a rigorous formulation of a quantum field theory. And some of the pitfalls he identified in the covariant approach in his 1963 lectures would then resurface as issues of the Wheeler-DeWitt equation, in particular the problem of operator ordering. For DeWitt the operator field equations appeared as functional derivatives of the effective action, rather than as canonical equations -- but the ambiguities from factor ordering were present in DeWitt's covariant action-based approach as well. For example, he highlighted (p. 181) that one might not be able to find an operator ordering that makes the field equations invariant with respect to q-number gauge transformations, i.e., to gauge transformations where the gauge parameter itself is a non-commuting quantum quantity.\footnote{Why we should want such an invariance in the first place is of course a different question.} He also argued that the correct operator ordering would be essential for obtaining a unitary S-matrix (pp. 191--193).

In general, the operator ordering ambiguity reflects an ambiguity in quantization as to the representation of polynomials in $p$ and $q$ owing from the fact that the Poisson bracket Lie algebra of all polynomials in $q$ and $p$ is not isomorphic to the commutator bracket Lie algebra of all polynomials in quantum operators $L\hat{q}$ and $L\hat{p}$ acting by left multiplication \citep{joseph:1970}. Operator ordering ambiguities can show up as differences in the eigenspectrum of the resulting operators as terms of the order of $O(\hbar)$. The particular problem relevant to the quantum representation of the Hamiltonian constraint in canonical gravity is ambiguity in the quantum representation of the term that takes the form $q^2 p^2$. DeWitt in fact studied this problem in details as far back as 1952 and argued based on spatial invariance that the preferred `Laplacian ordering' should be adopted such that (schematically) $\hat{p}\hat{q}^2 \hat{p}$. This ordering has the virtue that it leads to a self-adjoint operator. However, DeWitt, in writing down the Wheeler-DeWitt equation in 1967 adopted the ordering of the form $\hat{q}^2 \hat{p}^2$ on the basis that it allowed for the field variables taken at the same space-time point should to be regarded as freely commutable. Although this choice is part of the iconic statement of the equation \eqref{WdW} and is advertised by DeWitt as a major achievement of the paper, he also states in the 1967 paper that the ordering we choose doesn't actually matter.   

What did he make of these issues at in the 1963? In the Les Houches lectures he pointed out that issues of factor ordering ``cannot be fully discussed apart from the problem of renormalization.'' Let us look a bit closer at DeWitt's argument, and consider two terms (in the operator equations of motion, say) that differ by the exchange of two field operators, the simplest kind of factor ordering ambiguity. The difference between the two orderings (which is itself an operator) will be proportional to the commutator between the two exchanged operators. But since we assume all field theories to be local, this will be the commutator between two field operators evaluated at the same space-time point; it will thus be proportional to $\lim_{x\rightarrow y} D (x-y)$, where $D$ is a singular two-point function. This is divergent, so the difference between the orderings will be an operator with a divergent coefficient -- just like a counterterm in perturbative renormalization. DeWitt thus anticipated that factor ordering and renormalization would have to be treated together (and were in fact closely related). And though he did not offer any clear prescription, he anticipated that the renormalization one performed to get rid of perturbative divergences might actually be enough to resolve all operator ordering ambiguities and that thus ``the factor ordering problem in field theory may be completely trivial'' (p. 193).

As he would later write to Wheeler, DeWitt thus ultimately considered the problem of factor ordering ``a red herring whose only function is to delay real progress.''\footnote{Letter to John Wheeler of 23 October 1964. The letter is inserted in Wheeler's Relativity Notebook 13, between pages 60 and 61.} This assertion was based on the assumption that the ambiguities due to factor ordering could all be absorbed in the renormalization process. However, DeWitt was not able to prove this conjecture. In the 1967 paper, he introduced an explicit non-perturbative regularization and renormalization scheme, designed to remove the ambiguities from factor ordering; but he freely conceded that it was not clear whether this scheme would actually coincide with a perturbative scheme designed to removed ultraviolet divergences \citep[p. 1121]{DeWitt:1967}  -- if such a perturbative renormalization of quantum gravity was even possible! But for the time being this approach allowed DeWitt to bypass the problem of factor ordering and to choose the simplest ordering (all canonical momenta to the right) for convenience. Ultimately more crucial is the fifth problem of rigour, the definition of the inner product; for DeWitt, and even more so for John Wheeler, who considered this one of central problems in quantum gravity, as we shall discuss in the next section.

\subsection{Wheeler's Story}
\label{Sec:6}

We can begin John Wheeler's story at the same 1963 Les Houches summer school where DeWitt had lectured on the covariant quantization of non-abelian gauge theories and gravity. Wheeler gave a similarly expansive set of lectures, primarily on classical general relativity. In these lectures, he emphasized that one had only recently (he credited the 1960 senior thesis of his student David H. Sharp) fully understood the least-action formulation of general relativity. While the (Hilbert) action had been known for a long time, it had not been clear which data was to be fixed on the boundaries of the integration region (the analog of fixing the endpoints of the trajectory in Lagrangian particle mechanics). But now one understood that this data was the spatial metric on the spacelike hypersurfaces forming the boundary. 

Identifying these variables was also important, because in a quantum theory of gravitation these were the quantities on which the wave function would depend. In this context in particular (but also in general), Wheeler never spoke of the 3-metric, but rather of ``three-geometries,'' writing the wave function of quantum gravity as $\Psi(^{(3)}\mathcal{G})$. First, to emphasize the fact that two metrics related by a coordinate transformation did not really represent different boundary conditions; and second to indicate that one also had to fix non-metrical properties of the manifold, especially its topology. This was usually considered fixed, i.e., not subject to variation in the minimization of an action integral. But Wheeler envisaged quantum fluctuations of topology at the Planck scale, giving space-time a foam-like structure; consequently, topology was taken to be variable, just like the three-metric.

While making these very concrete statements about the mathematical structure and physical implications of quantum gravity, Wheeler also emphasized the foundational difficulties implied by such a theory. How, in particular, was the wave function $\Psi(^{(3)}\mathcal{G})$ to be normalized (p. 517)? Which configuration space should one integrate over when taking the inner product of the wave function with itself (or the inner product more generally)? One might think that this would simply be the ``superspace of all three-geometries.'' This would imply six degrees of freedom per point (the components of the three-metric) minus three spatial diffeomorphisms. But, as Wheeler emphasized, one of these degrees of freedom ``represents time; or more specifically [\ldots] it has to do with the choice of the spacelike hypersurface.'' This was the immediate stumbling block in constructing a quantum theory of gravity that Wheeler had identified, how to split ``$^{(3)}\mathcal{G}$ into the two parts,'' choice of hypersurface and the actual configuration space on that hypersurface.

In his notebooks, one can see that Wheeler thought about this problem as a lack of rigour (in our sense of physicists' rigour) in the theory of quantum gravity. On 9 May, 1964, Wheeler took notes on a conversation he had with Bruno Zumino and David Finkelstein on a British European Airways flight from Geneva to London. Wheeler said that his ``greatest puzzlement'' was how quantum theory would modify  the gravitational collapse of a massive star into what would soon come to be known as a black hole.\footnote{On Wheeler's gradual acceptance of the reality of full gravitational collapse and his adoption (and later promotion) of the concept of a black hole, see Furlan 2022 and 2024.} He was then challenged by Zumino: ``what was [Wheeler] waiting for, isn't gravitation theory all quantized [?]'' To which Wheeler replied:

\begin{quote}
I said so far as I understood one had so far eq[uatio]ns to solve but no solutions. [\ldots] I said I really didn't understand the business, most of all I did not understand which questions to ask (Wheeler Relativity Notebook 12, pp. 161--162).
\end{quote}

These are just the two aspects of lacking rigour that we have identified: the inability to do actual calculations and conceptual confusion. And the next entry in Wheeler's notebook makes it clear that it was indeed the problem with the inner product described above that was to blame for this state of affairs. Now in Cambridge on 10 May, Wheeler further reflected on the conversation of the day before, noting:

\begin{quote}
Issues about the q[uanti]z[atio]n of geometry. They go back to the idea that 3-geom[etry] carries info about time. From here to the issue I have troubled about so many times, how do \emph{normalization} integral -- over which part of function space to integrate? (Relativity Notebook 12, pp. 163)
\end{quote}

Significantly, it also appears that at this point, i.e., in the spring of 1964, Wheeler's thinking on quantum gravity was not yet connected to Peres's Hamilton-Jacobi work. Wheeler had always assumed that the three momentum constraints had to be solved before writing down the wave function of quantum gravity; as we have seen, he always wrote his state as a functional of 3-geometries, i.e., as a functional of the 3-metric but with the redundancy due to three-dimensional diffeomorphism invariance already somehow removed. In a notebook entry dated 29 May (Wheeler was still in Cambridge), we find reflections on ``Quantum Geometrodynamics'' (Notebook 12, pp. 209--211); here, Wheeler writes the normalization integral for the wave function as

\begin{equation}
1 = \int \left\vert \psi \left( ^{(3)}\underline{\mathcal{G}} \right) \right\vert^2 \; \; \mathcal{D}_2 \; ^{(3)}\underline{\mathcal{G}}  
\end{equation}

with the subscript 2 clearly indicating that one is integrating over only two degrees of freedom (per point); one is thus not only not integrating over variables associated with change in ``time'' (Wheeler's scare quotes), which Wheeler was so concerned with, but also not over the three additional variables associated with spatial diffeomorphisms. The underline in $^{(3)}\underline{\mathcal{G}}$ presumably indicates the removal of the temporal degree of freedom, which Wheeler considered to be a part of the three-geometry $^{(3)}\mathcal{G}$, as opposed to the degrees of freedom associated with spatial diffeomorphisms.

In an abortive attempt at quantizing the Hamiltonian constraint (i.e., in writing down something like the later Wheeler-DeWitt equation), Wheeler consequently represented the field momentum operators as

\begin{equation}
\underline{\pi} = \frac{\mathrm{const}}{i} \frac{\delta}{\delta ^{(3)}\underline{\mathcal{G}}}
\end{equation}

i.e., in a very formal way as functional derivatives with respect to three-geometries. Peres's Hamilton-Jacobi formulation suggested a different approach. He wrote the field momenta $\pi^{mn}$ as functional derivatives of his principal function $S$ with respect to the components of the three-metric $g_{mn}$; in the traditional Schrödinger quantization method, where $S$ becomes the phase of the wave function, this would imply that the field momentum operators are to be represented by functional derivatives with respect to the metric components -- a far more concrete (we might say: physically rigorous) representation. These differences in approach support our contention that the role of Peres was more marginal than that implied by Wheeler's recollections. For DeWitt the situation is somewhat more ambiguous: he cites Peres's work in his 1967 paper and he does represent the field momenta as functional derivatives with respect to the full three-metric -- this would be essential to his formulation of the inner product. The quantization method DeWitt presents in that paper, however, does not take off from the Hamilton-Jacobi equation for general relativity, but rather presents that equation quite late, as a derived result.

\subsection{The Role of the Inner-Product}

What, then, did Wheeler get from DeWitt? If our explanatory hypotheses is correct then one would expect that Wheeler had not got excited by the form of the equation at the airport meeting, but rather by the possibility of making it rigorous enough by the contemporary and contextual standards that he was adhering to, as well as enabling concrete calculations that would latch it onto empirical data. We are able to support this hypothesis via contemporary textual evidence.

Looking at Wheeler's notebooks, we find that DeWitt had sent Wheeler a long letter -- containing both the Wheeler-DeWitt equation and the scalar product -- on 19 June 1964, more than a month before the famous airport meeting. DeWitt presents his results as a response to Wheeler's ``query concerning the manifold for integration for 3-geometries,'' i.e., the problem highlighted in the previous section. And DeWitt's letter presents his formulation of the inner product, in particular, as a response to this question. This is emblematically captured in Wheeler's notes from the airport meeting, which took place on 8th August 1964 and was presumably convened as a reaction to DeWitt's letter.\footnote{This material is to be found in Wheeler's Relativity Notebook 13. The notes from the airport meeting are on pp. 9--10, while DeWitt's letter is inserted between pages 14 and 15.} The notes consist precisely of the inner product he was missing and not the form of the equation, which was more or less obvious:

\bigskip
\includegraphics[width=0.95\textwidth]{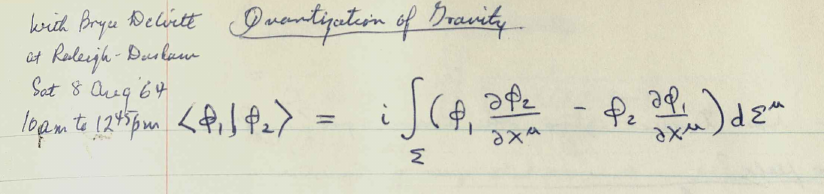}\label{wdwfig}

\noindent When Wheeler presented the new equation (perhaps for the first time) in October 1964, he characterized it as a ``wave equation related to gravity in same way as Klein-Gordon Eq. to physics of a single particle,'' thus emphasizing the analogy established through the inner product.\footnote{Notes taken by Dieter Brill on 20 October 1964.} The documentary evidence thus directly supports our explanatory hypothesis. Rigour matters in physics, even when it does not turn out to be magical.

In addition to DeWitt's evident concern with the factor ordering problem we also have contemporary evidence that DeWitt considered the inner product to be one of his key results. In particular, Don Salisbury \citep{salisbury:2022} has found a report to the Office of Naval Research that DeWitt wrote (in the 3rd person) for a grant that commenced in December, 1964, which runs as follows:
\begin{quote}
One of the senior investigators (B. S. D.) has recently proposed a functional equation for the state functional of the gravitational field in the so-called canonical
theory. He has proposed a functional integral expression for the inner product of two state vectors. This work has not yet been published but it has stimulated considerable activity both at North Carolina and at Princeton and has made it possible for the first time to discuss a number of fundamental issues in concrete form. Chief among these is the problem of gravitational collapse. In a reappraisal of the canonical quantum theory of gravity the attempt has been made
to find a dynamical interpretation of the so-called fourth Hamiltonian constraint. This has led to the discovery of
a six-dimensional diﬀerential hyperbolic manifold which underlies the intrinsic dynamics of the gravitational field
and the introduction of a functional diﬀerential `wave' equation of the second degree for the state functional of the
theory. This in turn has led to the discovery of a natural definition for the inner product of two state vectors.
\end{quote}
The obvious implication is that on DeWitt's view, if one was  without the functional integral expression for the inner product of two state vectors, then one would not be able to discuss a number of fundamental issues in canonical quantum gravity in concrete form. Thus questions of rigour and of physical application and conceptual clarity are understood to be closely related. 

Indeed, this analysis of what DeWitt took to be the substance of the `rigour problem' for the Wheeler-DeWitt equation is reflected in his focus within the paper on the physical motivation for and interpretation of the inner product. Let us take a somewhat closer look at DeWitt's inner product, as presented in his equation (5.19):

\begin{equation}
\label{DeWittIP}
(\Psi_b,\Psi_a)
= Z \int_{\Sigma} \Psi_b^{*}\!\left[^{(3)}\mathcal{G}\right]
\prod_{x}
\left(d\Sigma^{ij} G_{ijkl}\,
\frac{\overrightarrow{\delta}}{i\,\delta g_{kl}}
-
\frac{\overleftarrow{\delta}}{i\,\delta g_{kl}}\,
G_{ijkl}d\Sigma^{ij}
\right)
\Psi_a\!\left[^{(3)}\mathcal{G}\right].
\end{equation}

where the integration over $\Sigma$ is defined with respect to a $5 \times \infty^3$-dimensional surface in the $6 \times \infty^3$ space $\text{Riem }  \Sigma$, which is the space of Riemannian three-metrics $g_{ij}$ on $\Sigma$; $d\Sigma^{ij}$ are directed surface elements; and $Z$ is a normalization constant. The analogy with the Klein-Gordon inner product product can be recognised if we consider the standard form induced by the conserved current:
\begin{equation}
(\phi_1,\phi_2)_{KG}
=
i \int_{\Sigma}
\left(
\phi_1^{*}\,(\partial_{\mu}\phi_2 )
-
(\partial_{\mu}\phi_1^{*})\,\phi_2
\right)
\, d\Sigma^{\mu},
\end{equation}
where $d\Sigma^{\mu}=n^{\mu}d\Sigma$ is the directed hypersurface element. How does this inner product now resolve Wheeler's problem of the degree of freedom associated with temporal change? DeWitt's central insight was that one has a similar problem already in special-relativistic quantum mechanics: the Klein-Gordon wave function is a function of space and time on equal, covariant footing. However, in order to establish some analogy with non-relativistic quantum mechanics (e.g., to calculate transition probabilities or apply the Born rule), one had to construct an inner product that was both Lorentz-invariant \emph{and} only involved integration over a spacelike three-surface.\footnote{The construction starts from the normalization integral, which one obtains by first building from the wave function a covariant four-current and then integrating the orthogonal component of that current (the ``density'') over the three-surface. The scalar product is then constructed by analogy.} DeWitt then generalised this construction principle to the far more complex gravitational case.

What is important is \textit{not} that DeWitt's Klein-Gordon inspired `inner product' provides a mathematically rigorous  object that supplies the relevant norm structure for the physical Hilbert space of canonical quantum gravity or even a finite probability measure. It doesn't. Indeed, the expression \eqref{DeWittIP} induces a `measure' that is neither finite, nor positive, nor capable of supporting a well-defined $L^2$ space.\footnote{The inner-product issue is closely related to the problem of time. For further discussion see \cite[p. 151]{Kiefer:2012}, \cite{Kuchar:1991,Kuchar:1992} and \cite[5.2.2]{Isham:1992}.} It is this application that the rigorisation of the inner product associated with the Wheeler-DeWitt equation \eqref{WdW} is targeted towards, not the exogenous problem of mathematical rigour.

\section{Concluding Remarks}
\label{sec:7}

We offered a plausible explanatory hypothesis that it was the expression for the inner-product that was crucial in establishing the Wheeler-DeWitt form of the equation that bears their name. This is what meant that the equation was \textit{rigorous enough by the standards of the day} to be accepted as well-formed and at least part of the reason why Wheeler and DeWitt got credit. This was a standard of rigour according to working theoretical physicists -- and it was not parasitic on mathematicians rigour in any clear sense. It is a separate norm of mathematical discourse that is highly local within a community of theoretical physicists working on a particular problem set. 

\section*{Acknowledgements}

This work was supported by the Benjamin Meaker Fellowship programme at the University of Bristol. Thanks to Kurt Sundermeyer, Rami Jreige, Claus Kiefer and Domenico Giulini for helpful comments on a draft manuscript and the audiences in Schloss Ringberg and Paris for feedback.  

 \bibliographystyle{chicago}
\bibliography{Masterbib3,refs}

\end{document}